\documentclass[fleqn,twoside]{article}
\usepackage{espcrc2}

\usepackage{graphicx}
\usepackage[figuresright]{rotating}

\newcommand{\AmS}{{\protect\the\textfont2
  A\kern-.1667em\lower.5ex\hbox{M}\kern-.125emS}}

\title{Core-collapse supernovae and gravitational waves}

\author{Christian Y. Cardall\address{Physics Division, Oak Ridge National Laboratory, Oak Ridge,
 	TN 37831-6354}%
      \address{Department of Physics and Astronomy, University of Tennessee,
	Knoxville, TN 37996-1200}%
        \thanks{This work was supported 
by  Scientific Discovery Through
Advanced Computing (SciDAC), a program of the Office of Science of the U.S. Department of Energy (DoE); and by Oak Ridge National Laboratory, managed by UT-Battelle, LLC, for the DoE under contract DE-AC05-00OR22725.}%
}
       
\begin{document}

\begin{abstract}
Core-collapse supernovae are dramatic events with a rich phenomenology, including gravitational radiation. Simulations of these events in multiple spatial dimensions with 
energy- and angle-dependent neutrino transport are still in their infancy.
Core collapse and bounce in a supernova in our galaxy may well be visible by 
first-generation LIGO, but detailed understanding waits on improvements in 
modeling both stellar progenitors and the supernova process. 
\vspace{1pc}
\end{abstract}

\maketitle

\section{CORE-COLLAPSE SUPERNOVAE}

Core-collapse supernovae---those 
of Type Ib, Ic, and II---result from the catastrophic collapse of the core of a massive star. Depending on the properties of the progenitor star, the collapse may result in a 
black hole; alternatively, the remnant can be a neutron star with 
$GM/R \sim 0.1 c^2$, where $G$ is the 
gravitational constant, $M$ and $R$ are the neutron star mass and radius, and
$c$ is the speed of light. Such a system is manifestly relativistic; and if the
high densities and infall velocities implied by the above relation are combined
with sufficient asphericity, the violence of core collapse and its aftermath may
be expected to produce significant gravitational radiation. 
(Type Ia supernovae, which are caused by the thermonuclear detonation or
deflagration of a white dwarf star, are not expected to be interesting sources of
gravitational radiation.)

I begin by outlining our current understanding (and lack of 
understanding) of the core-collapse supernova process.

For most of their existence stars burn hydrogen into helium. 
In stars at least eight times as massive as the Sun, 
temperatures and densities become sufficiently high to burn 
to carbon, oxygen, neon, magnesium, and silicon and iron group elements. 
The iron group nuclei are the most tightly bound, and here burning in 
the core ceases. 

The iron core---supported by electron 
degeneracy pressure---eventually 
becomes unstable. Its inner portion undergoes homologous collapse
(velocity proportional to radius), and the outer portion collapses 
supersonically. 
Electron capture on nuclei is one instability leading to collapse, and 
this process continues throughout collapse, producing neutrinos.
These neutrinos escape freely
until densities and temperatures in the collapsing core
become so high that even neutrinos are 
trapped. 

Collapse is halted soon after the matter exceeds nuclear density; 
at this point (called ``bounce''), a shock wave forms at the boundary between the homologous
and supersonically collapsing regions. The shock begins to move out,
but after the shock passes
some distance beyond the surface of the newly born neutron star, 
it stalls as energy 
is lost to neutrino emission and dissociation of heavy 
nuclei falling through the shock.

The details of how the stalled shock is revived
sufficiently to continue plowing through the outer layers of the
progenitor star are unclear. Some combination of neutrino heating of
material behind the shock, convection, instability of the spherical
accretion shock, rotation, and magnetic fields launches the explosion.

It is natural to consider neutrino heating as a mechanism for
shock revival, because neutrinos dominate the energetics of
the post-bounce evolution.
Initially, the nascent neutron star is a hot thermal bath of dense nuclear matter, 
electron/positron pairs, photons, and neutrinos, containing most of 
the gravitational potential energy released during core collapse. 
Neutrinos, having the weakest interactions, are the most efficient 
means of cooling; they diffuse outward on a time scale of seconds, 
and eventually escape with about 99\% of the released gravitational energy.

Because neutrinos dominate the energetics of the system, 
a detailed understanding of their evolution will be integral to any
detailed and definitive account of the supernova process.
If we want to understand the explosion---which accounts for only about 1\% of 
the energy budget of the system---we should carefully account for the
neutrinos' much larger contribution to the energy budget.

What sort of computation is needed to follow the neutrinos' evolution?
 Deep inside the newly-born neutron star, 
 the neutrinos and the fluid are tightly coupled (nearly in equilibrium);  but as the neutrinos are transported from inside the neutron star, they go from a nearly isotropic diffusive regime to a strongly forward-peaked free-streaming region. Heating of material behind the shock occurs precisely in 
this transition region, and modeling this process accurately requires tracking both the energy and angle dependence of the neutrino distribution functions at every point
in space. 

While a full treatment of this six-dimensional neutrino radiation hydrodynamics
problem remains too costly for currently available computational resources, there is much that has been 
learned over the years through detailed modeling.

\section{SIMULATING THE EXPLOSION}

Supernovae have a rich phenomenology---observations of many types that 
modelers would like to reproduce and explain. 
Chief among these is the explosion itself, which is not yet produced 
robustly and convincingly in simulations. 
Other observables of interest include neutrino signatures; 
neutron star spins,
kick velocities, and magnetic fields; synthesized element
abundances; all kinds of measurements across the electromagnetic
spectrum; and of course the subject of this conference session, gravitational waves. 

Simulations of core collapse, bounce, and its immediate aftermath
have mostly aimed at the first few of these observables: 
the explosion mechanism, neutrino signatures, remnant pulsar
properties, and gravitational waves. 
I will now describe some of the progress in this work in
the past decade or so, focusing in particular on the explosion mechanism.

Throughout the 1990s, several groups performed simulations
in two spatial dimensions. Even in two spatial dimensions, 
computational limitations required approximations that
simplified the neutrino
transport. 

One simplification allowed for neutrino transport in two 
spatial dimensions, but with
neutrino energy and angle dependence integrated out---effectively
reducing a five dimensional problem to a two dimensional one
(see for example \cite{hera94,burr95}).  
These simulations exhibited explosions, suggesting that the enhancements 
in neutrino heating behind the shock resulting from convection provided a
robust explosion mechanism. More recent simulations in three spatial
dimensions with this same
approximate treatment of neutrino transport showed similar outcomes
\cite{fryer02}.

A different simplification of neutrino transport employed in the 1990s
was the imposition of energy-dependent
neutrino distributions from spherically symmetric simulations
onto fluid dynamics computations in two spatial dimensions
\cite{mezz98b}. Unlike the multidimensional simulations
discussed above, these did not exhibit explosions, casting doubt
upon claims that convection-aided
neutrino heating constituted a robust explosion mechanism.

The nagging qualitative difference between multidimensional
simulations with different neutrino transport approximations 
renewed the motivation for simulations in which both the energy
and angle dependence of the neutrino distributions were retained.
Of necessity, the first such simulations were performed in
spherical symmetry (actually a three-dimensional problem, depending
on one space and two momentum space variables). Results from
three different groups are in accord: Spherically symmetric models
do not explode, even with solid neutrino transport 
\cite{buras02,thompson02,liebendoerfer02}. 

Recently, one of these groups performed simulations in two spatial
dimensions, in which
their energy- and angle-dependent neutrino transport was made partially
dependent on spatial polar angle as well as radius \cite{janka02,buras03}.
Explosions were not seen in any of these simulations, except for one in which
certain terms in the neutrino transport equation corresponding to Doppler shifts
and angular aberration due to fluid motion were dropped. This was a surprising
qualitative difference induced by terms contributing what are typically thought of as small corrections. The continuing lesson
is that getting the details of the neutrino transport right makes a difference. 

Where, then, do simulations aiming at the explosion mechanism stand?
The above history suggests that elucidation of the mechanism will
require simulations that feature truly spatially multidimensional 
neutrino transport. In addition,
inclusion of magnetic field dynamics seems increasingly strongly motivated
as a possible driver of the explosion,
because simulations with ``better'' neutrino transport have failed to 
explode---even in multiple spatial dimensions.

\section{GRAVITATIONAL RADIATION}

Core bounce and associated phenomena 
(halt of the collapse, shock formation, and ``ring-down'')
would be the strongest source of gravitational radiation from a core-collapse
supernova. Should such an event occur in our galaxy (this happens only once
every $\sim$few-several decades), it probably would be detectable 
by first-generation gravitational wave interferometers.
This conclusion is reached in several studies; a recent example is the
work of Ott et al. \cite{ott03}. 

These authors also make some points that highlight
the need for more complete simulations. Most often, the results of
spherically symmetric stellar evolution computations are artificially put
into rotation for use as progenitors in studies of core collapse;
but in simulations with the most recent progenitors that include rotation and magnetic
fields, the gravitational radiation can be an order of magnitude weaker.
In addition, some features of the waveform (such as coherent post-bounce
oscillations) probably will change with the inclusion of realistic neutrino transport because
of the associated stalling of the shock.

In addition to the violence of core bounce and its immediate
aftermath,
additional phenomena in the subsequent evolution may produce
gravitational radiation at lower amplitudes. These include triaxial instabilities in the
neutron star, anisotropic neutrino emission, convection, and
(on longer time scales than the supernova process itself) R-mode 
instabilities \cite{talks}. Another less well-known possible source of gravitational
waves is the standing accretion shock
instability \cite{blondin02}. 

\end{document}